\documentclass[12pt,a4paper]{article}
\usepackage{a4wide}
\usepackage{latexsym}
\usepackage{epsf}
\usepackage{amssymb}
\usepackage{graphics}

\begin{document}
\def\be{\begin{equation}}
\def\ee{\end{equation}}
\def\bea{\begin{eqnarray}}
\def\eea{\end{eqnarray}}

\def\pd{\partial}
\def\a{\alpha}
\def\b{\beta}
\def\bi{\begin{itemize}}
\def\ei{\end{itemize}}
\def\g{\gamma}
\def\d{\delta}
\def\m{\mu}
\def\n{\nu}
\def \h{\mathcal{H}}
\def \hh{\mathcal{G}}
\def\t{\tau}
\def\p{\pi}
\def\th{\theta}
\def\l{\lambda}
\def\O{\Omega}
\def\r{\rho}
\def\cd{\cos{y_2}}
\def\cu{\cos{y_1}}
\def\su{\sin{y_1}}
\def\sd{\cos{y_2}}

\def\s{\sigma}
\def\e{\epsilon}
  \def\scri{\mathcal{J}}
\def\cM{\mathcal{M}}
\def\tcM{\tilde{\mathcal{M}}}
\def\RR{\mathbb{R}}

\hyphenation{re-pa-ra-me-tri-za-tion}
\hyphenation{trans-for-ma-tions}


\begin{flushright}
IFT-UAM/CSIC-04-26\\
gr-qc/0405107\\
\end{flushright}

\vspace{1cm}

\begin{center}

{\bf\Large   Quantum Gravity}

\vspace{.5cm}

{\bf Enrique \'Alvarez }

\vspace{.3cm}

\vskip 0.4cm  
 
{\it  Instituto de F\'{\i}sica Te\'orica UAM/CSIC, C-XVI,
and  Departamento de F\'{\i}sica Te\'orica, C-XI,\\
  Universidad Aut\'onoma de Madrid 
  E-28049-Madrid, Spain }\footnote{Lectures given in the 40th Karpacz Winter School.}

\vskip 0.2cm

\vskip 1cm

{\bf Abstract}

General lectures on quantum gravity.

\end{center}

\author{Enrique Alvarez}
%
%


\section{General Questions on Quantum Gravity}
It is not clear at all what is the problem in quantum gravity (cf. \cite{Alvarez} or 
\cite{Alvarez-Gaume} \index{Quantum Gravity} for 
general reviews, written in the same spirit as the present one).
The answers to the following questions are not known, and I believe it can do no harm
to think about them before embarking in a more technical discussion.
\par
To begin, it has been   proposed that gravity should not be quantized, owing to its
special properties as determining the background on which all other fields propagate.
There is a whole line of thought on the possibility that gravity is not a fundamental theory,
and this is certainly an alternative  one has to bear in mind. Indeed, even the {\em holographic
principle} of G. 't Hooft, to be discussed later, can be interpreted in this sense.
\par
Granting that, the next question is whether 
it does  make any sense to consider gravitons
  propagating in some background; that is, whether there is some useful
approximation in which there is a particle physics approach to the physics
of gravitons as quanta of the gravitational field. A related question is whether 
semiclassical gravity, i.e., the approximation
in which the source of the classical Einstein equations is replaced by the
expectation value of the energy momentum tensor of some quantum theory has some physical 
(\cite{Duff}) validity in some limit. We shall say more on this problems towards the end.
\par
At any rate, even if it is possible at all, the at first sight easy problem of graviton
 interactions
in an otherwise flat background has withstood analysis of several generations of physicists.
The reason is that the coupling constant has mass dimension $-1$, so that the structure
of the perturbative counterterms involve higher and higher orders in the curvature invariants
(powers of the Riemann tensor in all possible independent contractions), schematically,
\be
S=\frac{1}{2\kappa_R^2}\int R+\int R^2 + \kappa_R^2 \int R^4+\ldots
\ee
Nobody knows how to make sense of this approach, except in one case, to be mentioned later on.
\par
It could be possible, {\em sensu stricto} to stop here. But if we believe that 
quantum gravity should  give answers to such questions as to the fate of the initial 
cosmological singularity, its is almost unavoidable to speak of the wave function of the 
universe.
This brings its own set of problems, such as to whether it is possible to do quantum mechanics
 without classical observers or whether  the wave function of the Universe
 has a probablilistic interpretation. Paraphrasing C. Isham \cite{Isham}, one would 
not known when to 
qualify a probabilistic prediction on the whole Universe as a successful one.
\par
The aim of the present paper is to discuss in some detail established results on the field.
In some strong sense, the review could be finished at once, because there are none. 
There are, nevertheless, some interesting attempts, which look promising from certain 
points of view.
Perhaps the two approaches that have attracted more attention have been the loop approach,
on the one hand and strings on the other. We shall try to critically assess prospects in both.
 Interesting related papers are \cite{Horowitz}\cite{Smolin}.
\par
Even if for the time being there is not (by far) consensus on the scientific community
of any quantum gravity physical picture, many great physicist have not been able to resist the
 temptation of working (usually only for a while) on it. This has produced a huge spinoff in
quantum field theory; to name only a few, constrained quantization, compensating ghosts,
background field expansion and topological theories are concepts or techniques first 
developped in thinking about these problems, and associated to the names of Dirac, Pauli,
Weinberg, Feynman, deWitt, Witten etc. In many cases, more or less surprising relationships
have been found with quantum gauge theories. There are probably more in store, if one is to
judge from the success of the partial implementation of holographic ideas in 
Maldacena's conjecture (more on this later).
\par
This should be kept into account when reading the references. We have not atempted to be
comprehensive, and we have used only the references familiar to us; but in some of the
references, in particular in our own review article of 1989 (\cite{Alvarez}
 there are more entry points
into the vast literature. After all, paraphrasing Feynman, we still do not know what could
be relevant in a field until the main problems are solved.

\section{The issue of background independence}
One of the main differences between both attacks to the quantum gravity problem
 is the issue of background
 independence, by which it is understood that no particular background should enter
into the definition of the theory itself. Any other approach is purportedly at
variance with diffeomorphism invariance.
\par
Work in particle physics in the second half of last century led to some
understanding of ordinary gauge theories. Can we draw some lessons from there?
\par
Gauge theories can be formulated in the {\em bakground field approach}, as
introduced by B. de Witt and others (cf. \cite{Dewitt}). In this approach,
the quantum field theory depends on a background field, but not on any one in particular,
and the theory enjoys background gauge invariance.
\par
Is it enough to have quantum gravity formulated in such a way? \footnote{This was, 
incidentally, the way
G.  Hooft and M. Veltman did the first complete one-loop calculation (\cite{thv}).}
\par
It can be argued that the only 
vacuum expectation value consistent with diffeomorphisms invariance is
\be
<0|g_{\a\b}|0>=0
\ee
in which case the answer to the above question ought to be in the negative, because 
this is a singular
background and curvature invariants do not make sense.
It all boils down as to whether the ground state of the theory is diffeomorphism
 invariant or not.
There is an example, namely three-dimensional gravity in which invariant quantization 
can be performed
\cite{Witten3}. In this case at least, the ensuing theory is almost topological. 
\par
In all attempts of a canonical quantization of the gravitational field, 
one always ends up with an (constraint) equation
corresponding physically to the fact that the total hamiltonian of a parametrization 
invariant theory
should vanish. When expressed in the Schr\"odinger picture, this equation is often 
dubbed the {\em Wheeler-de Witt equation}. This equation is plagued by
operator ordering and all other sorts of ambiguities.
It is curious to notice that in ordinary quantum field theory there also exists a Schr\"odinger
representation, which came recently to be controlled well enough as to be able to 
perform lattice 
computations (\cite{Luscher}).
\par
Gauge theories can be expressed in terms of gauge invariant operators, such as
Wilson loops . They obey a complicated
set of equations, the loop equations, which close in the large $N$ limit
as has been shown by Makeenko and Migdal (\cite{MM}). These equations can be properly 
regularized,
e.g. in the lattice. Their explicit solution
is one of the outstanding challenges in theoretical physics. Although many conjectures
have been advanced in this direction, no definitive result is available.

 \section{ The canonical approach.}

It is widely acknowledged that there is a certain tension between a $(3+1)$ decomposition
implicit in any canonical approach, privileging a particular notion of time, and the
 beautiful geometrical structure of general relativity, with its invariance under general
coordinate transformations. 
\par
Let us now nevertheless explore how far we can go on this road, following the still very 
much worth
reading work of deWitt (\cite{Dewitt}).
\par
If we are given a spacelike surface (which will represent physically all spacetime events
to which it will be asigned a fixed time),  say
\be
y^{\a}=f^{\a}(x^i)\nonumber
\ee

\begin{figure}[!ht] 
\begin{center} 
\leavevmode 
\epsfxsize= 10cm

\epsffile{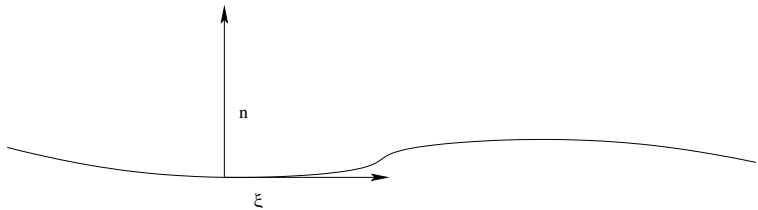} 
\caption{Spacelike surface of codimension one.}
\label{fig:1}   
\end{center} 
\end{figure}

The tangent vectors to the surface are
\be
\xi_i\equiv\pd_i f^{\a}\pd_{\a}\nonumber
\ee 
and the induced metric (that is, the pull-back to the surface of the spacetime metric) is
\be
h_{ij}\equiv g_{\a\b}\xi_i^{\a}\xi_j^{\b}\nonumber
\ee 

The unit normal is then defined as; 
\bea
&&g_{\a\b}n^{\a}\xi_i^{\b}=0\nonumber\\
&&n^2\equiv g_{\a\b}n^{\a}n^{\b}=1\nonumber
\eea

We are interested now in a set of such surfaces which covers all spacetime; that is, a 
foliation of (a portion of) the spacetime; namely a one-parameter
family of spacelike disjoint hypersurfaces
 \be
\Sigma_t\equiv\{y^{\a}= f^{\a}(x^i,t)\}\nonumber
\ee 

\begin{figure}[!ht] 
\begin{center} 
\leavevmode 
\epsfxsize= 10cm

\epsffile{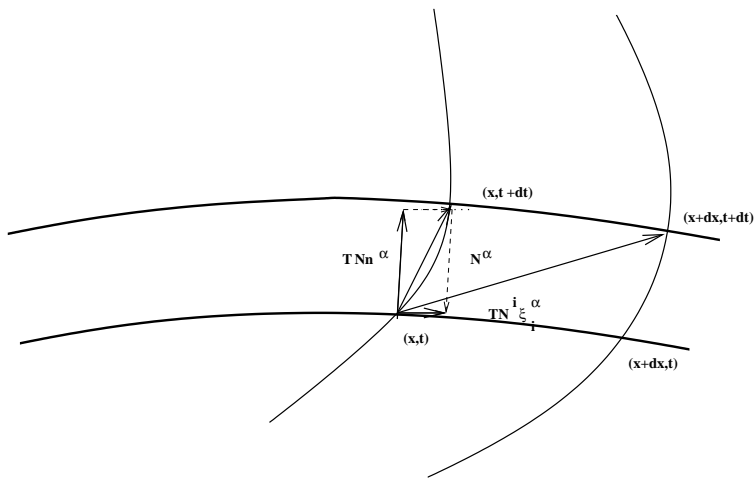} 
\caption{ADM lapse and shift variables.}
\label{fig:2}   
\end{center} 
\end{figure}

In a classical analysis Arnowitt, Deser and Misner (ADM)(\cite{ADM})
\index{ADM} characterized the embedding via two functions: the {\em lapse} and the {\em shift} :
we first define the vector
\be
N^{\a}\equiv \frac{\pd f^{\a}}{\pd t}\nonumber
\ee
in terms of which the lapse, $N$, is just the projection in the direction of the normal, and the
 {\em shift}, $N_i$ the (three) projections tangent to the hypersurface.
\be
N^{\a}= N n^{\a} + N^{i}\xi_i^{\a}\nonumber
\ee

All this amounts to a particular splitting of the full spacetime metric:
\be
ds^2 = g_{\a\b}dx^{\a} dx^{\b}= g_{\a\b}df^{\a} df^{\b}=
g_{\a\b}(N^{\a}dt+\xi_i^{\a}dx^i)(N^{\b}dt+\xi_j^{\b}dx^j)
= N^2 dt^2 + h_{jk}(N^j dt+ dx^j)(N^k dt +dx^k)\nonumber
\ee
or, what is the same,
\be
g_{\m\n}=h^{ij}\xi_{i\m}\xi_{j\n}+n_{\m}n_{\n}\nonumber
\ee

All surfaces which are equivalent from the intrinsic point of view, can be however embedded
differently; the extrinsic curvature discriminates between them:
\be
K_{ij}=-\xi^{\a}_i\nabla_{\rho}n_{\a}\xi^{\rho}_j\nonumber
\ee

The Gauss- Codazzi equations relate intrinsic curvatures associated with the intrinsic
geometry in the hypersurface with spacetime curvatures
precisely through the extrinsic curvature:
\be
R[h]_{lijk}=R[g]_{\a\b\sigma\rho}\xi^{\a}_l\xi^{\b}_i\xi^{\sigma}_j\xi^{\rho}_k-
K_{ij}K_{lk}+K_{ik}K_{lj}
\ee
and
\be
\nabla[h]_k K_{ij}-\nabla[h]_j K_{ik}=
R_{\a\b\sigma\rho}n^{\a}\xi^{\b}_i\xi^{\rho}_j\xi^{\sigma}_k
\ee

whereas the curvature scalar is given by
\be
R=R^{\a\b}\,_{\a\b}=2 R^{ni}\,_{ni}+R^{ij}\,_{ij}
\ee
In terms of this splitting, the  Einstein-Hilbert action reads:
\be
L_{EH}\equiv \sqrt{g}R[g]=N\sqrt{h}(R[h]+K_{ij}K^{ij}-K^2)-\pd_{\a}V^{\a}
\ee
with
\be
V^{\a}=2\sqrt{g}(n^{\b}\nabla_{\b}n^{\a}-n^{\a}\nabla_{\b}n^{\b})\nonumber
\ee

Primary constraints appear when defining canonical momenta:

\be
p^{\m}\equiv\frac{\pd L}{\pd\dot{N}_{\m}}\sim 0\nonumber
\ee
the momenta conjugate to the spatial part of the metric is:
\be
\pi^{ij}\equiv\frac{\d L}{\d \dot{h}_{ij}}=-h^{1/2}(K^{ij}-K h^{ij})\nonumber
\ee
The canonical conmutation relations yield:
\be
\{\pi^{ij}(\vec{x}),h_{kl}(\vec{x}\,^{\prime})\}=
-\d(\vec{x},\vec{x}\,^{\prime})\frac{1}{2}(\d^i_k\d^j_l+\d^j_k\d^i_l)\nonumber
\ee

The total Hamiltonian reads

\be
H\equiv\int d^3 x (\pi_{\m}\dot{N}^{\m}+\pi^{ij}\dot{h}_{ij}-L)=
\int d^3 x(N{\cal H}+N^i{\cal H}^i)\nonumber
\ee  
where
\be
{\cal H}(h,\pi)=h^{-1/2}(\pi_{ij}\pi^{ij}-\frac{1}{2}\pi^2)-h^{1/2} R[h]
\ee

and

\be
 {\cal H}_i(h,\pi) =-2 h_{ik}\pd_j \pi^{kj}-(2\pd_jh_{ki}-
\pd_ih_{kj})\pi^{kj}=-2 \nabla[h]_j\pi_i\,^j 
\ee

The system of constraints is now consistent  (that is, that the classical time evolution 
of the  constraints
is still a linear combination of constraints): 
\be
\dot{p}^{\m}=\{p^{\m},H\}=({\cal H},{\cal H}_i)\sim 0\nonumber
\ee
Second class constraints 
\be
N^{\m}=f^{\m}\nonumber
\ee
can now be imposed. The whole hamiltonian analysis boils down to the two constraint equations
\bea
{\cal H}=0\nonumber\\
{\cal H}_i=0\nonumber
\eea
Much of the preceding analysis is actually quite generic for generally covariant systems. 
The full set of constraints obeys the  Dirac-Schwinger algebra
\bea
&&\{{\cal H}(\vec{x}),{\cal H}(\vec{y})\}=[{\cal H}^i(\vec{x})+
{\cal H}^i(\vec{y})]\pd_i\d(\vec{x},\vec{y})\nonumber\\
&&\{{\cal H}_i(\vec{x}),{\cal H}(\vec{y})\}={\cal H}(\vec{x})\pd_i\d(\vec{x},\vec{y})\nonumber\\
&&\{{\cal H}_i(\vec{x}),{\cal H}_j(\vec{y})\}={\cal H}_i(\vec{y})\pd_j \d(\vec{x},\vec{y})+
{\cal H}_j(\vec{x})\pd_i\d(\vec{x},\vec{y})
\eea
which is nothing else than  the $\Sigma$-projected algebra of the Diff(M) group.

\par

Usually  no reduction is made on the dynamical variables of the system, which amounts to keep 
$h_{ij},\pi^{ij}$ as (redundant) quantum variables.
It is not clear how  singular metrics can be avoided, because it is not easy to impose the
condition that the metric is a positive definite operator.
\par
Physical states in the Hilbert space are provisionally defined \`a la Dirac
\bea
&&{\hat{{\cal H}}}|\psi>=0 \nonumber\\
&&{\hat{{\cal H}}}_i |\psi>=0 \nonumber
\eea

It has been realized since long that this whole  approach suffers from the  {\em frozen time
problem}, i.e., the Hamiltonian reads

\be
H\equiv \int d^3 x(N{\cal H}+N^i{\cal H}^i)\nonumber
\ee
so that acting on physical states
\be
{\hat H}|\psi>=0
\ee

in such a way that Schr\"odinger's equation 

\be
i\frac{\pd}{\pd t}|\psi>={\hat{H}}|\psi>
\ee
seemingly forbids any time dependence.

\par
There are many unsolved problems in this approach, which has been kept at a formal level.
The first one is an obvious operator ordering ambiguity owing to the nonlinearity.
In the same vein, it is not clear whether it is possible to make the constraints hermitian.
Besides, it is not clear that one recovers the full Diff invariance from the
Dirac- Schwinger algebra. Actually, it is not known whether this is necessary; that is, 
what is the full symmetry of the quantum theory.
\par

We can proceed further, still formally\footnote{ It is bound to be formal as long as
the problem of the infinities is not fully addressed. We know from the analysis of this
representation for gauge theories in the lattice that those are the most difficult 
problems to solve.}, using the Schr\"odinger representation
defined in such a way that
\be
(\hat{h}_{ij}\psi)[h]\equiv h_{ij}(x)\psi [h]
\ee
and
\be
(\hat{\pi}^{ij}\psi)[h]\equiv - i \hbar \frac{\d\psi}{\d h_{ij}(x)}[h]
\ee
If we assume that diffeomorphisms act on wave functionals as:
\be
\psi[f^{*}h]=\psi[h]
\ee
then the whole setup for the quantum dynamics of the gravitational field lies in
Wheeler's {\em superspace} (nothing to do with supersymmetry) which is
the set of three-dimensional metrics modulo three-dimensional diffs : 
$Riem(\Sigma)/Diff(\Sigma)$.
\par

The Hamiltonian constraint then implies the famous {\em Wheeler-de Witt} equation.

\be
-\hbar^2 2\kappa^2 G_{ijkl}\frac{\d^2 \psi}{\d h_{ik}\d h_{jl}}[h]
-\frac{h}{2\kappa^2} R^{(3)}[h]\psi[h]=0
\ee
where the de Witt tensor is:
\be
G_{ijkl}\equiv\frac{1}{\sqrt{h}}\bigg(h_{ij}h_{kl}-\frac{1}{2}h_{ik}h_{jl}\bigg)
\ee
Needless to say, this equation, suggestive as it is,  is plagued with ambiguities.
The manifold of positive definite metrics has been studied by deWitt. He showed that it has 
signature $(-1,+1^5)$, where the timelike coordinate is given by the breathing mode
of the metric:
\be
\zeta=\sqrt{\frac{32}{3}}h^{1/4}
\ee
and in terms of other five coordinates $\zeta^a$ orthogonal to the timelike coordinate, 
the full metric reads
\be
ds^2=-d\zeta^2 + \frac{3}{32}\zeta^2 g_{ab}d\zeta^a d\zeta^b
\ee

with
\be
g_{ab}= tr\, h^{-1}\pd_a h h^{-1}\pd h
\ee
The  five dimensional submanifold with metric $g_{ab}$ is the coset space
\be
SL(3,\mathbb{R})/SO(3)
\ee
It has been much speculated whether the timelike character of the dilatations lies at the root
of the concept of time. The Wheeler-deWitt equation can be written in a form quite similar to
the Klein-Gordon equation:
\be
\bigg(-\frac{\pd^2}{\pd\zeta^2}+\frac{32}{3 \zeta^2}g^{ab}\pd_a\pd_b+\frac{3}{32}\zeta^2 R^{(3)}
\bigg)\Psi=0
\ee

The analogy goes further in the sense that also here there is a naturally defined scalar product
which is not positive definite:
\be
(\psi,\chi)\equiv\int_{\Sigma}\psi^{*}d\Sigma^{ij}G_{ijkl}\frac{\d \chi}{i\d h_{kl}}
-\chi^{*}d\Sigma^{ij}G_{ijkl}\frac{\d \psi}{i\d h_{kl}}
\ee

\section{ Using Ashtekar and related variables}
The whole philosophy of this approach is canonical, i.e., an analysis of the evolution of
variables defined classically through a foliation of spacetime by a family of spacelike 
three-surfaces
$\Sigma_t$. The standard choice in this case as we have just reviewed, is the 
three-dimensional metric, $g_{ij}$, and its canonical conjugate, related to the extrinsic 
curvature.
\par

Here, as in any canonical approach the way one chooses the canonical variables is fundamental.
\par
Ashtekar's clever insight started from the definition of an original set of 
variables (\cite{Ashtekar}) stemming from the Einstein-Hilbert 
lagrangian written in the form \footnote{Boundary terms have to be considered as well. 
We refer to 
the references for
details.}
 \be
S=\int e^a\wedge e^b\wedge R^{cd}\epsilon_{abcd}  
\ee
where $e^a$ are the one-forms associated to the tetrad,
\be
e^a\equiv e^a_{\m}dx^{\m}.
\ee
Tetrads are defined up to a local Lorentz transformation
\be
(e^a)^{\prime}\equiv L^a\,_b(x)e^b
\ee
The associated $SO(1,3)$ connection one-form $\omega^a\,_b$ is usually called the 
{\em spin connection}. Its field strength is the  curvature expressed as a two form:
\be
R^a\,_b\equiv d\omega^a\,_b+\omega^a\,_c\wedge \omega^c\,_b.
\ee
Ashtekar's variables are actually based on the $SU(2)$ self-dual connection
 \be
 A=\omega - i * \omega
\ee
Its field strength is
\be
F\equiv d A + A\wedge A
\ee
The dynamical variables are then $(A_i, E^i\equiv F^{0i})$. The main virtue of these
 variables is that
constraints are then linearized.
One of them is exactly analogous to            
Gauss'law: 
\be
D_i E^i=0.
\ee
There is another one related to three-dimensional diffeomorphisms invariance,
\be
tr\, F_{ij}E^i=0
\ee
 and, finally, there is the Hamiltonian constraint,
\be 
 tr F_{ij}E^i E^j=0
\ee
\par
On a purely mathematical basis, there is no doubt that Astekhar's variables are of a 
great ingenuity.
As a physical tool to describe the metric of space, they are not real in general. This forces
a reality condition to be imposed, which is akward. For this reason it is usually prefered
to use the Barbero-Immirzi (\cite{Barbero}\cite{Immirzi}) 
formalism in which the connexion depends on a free parameter, $\gamma$,
\be
A_a^i=\omega_a^i +\gamma K_a^i
\ee
$\omega$ being the spin connection and $K$ the extrinsic curvature. When $\gamma=i$ Astekhar's
formalism is recovered; for other values of $\gamma$ the explicit form of the constraints 
is more complicated. Thiemann (\cite{Thiemann}) has proposed a form for the 
Hamiltonian constraint
which seems promising, although it is not clear whether the quantum constraint 
algebra is isomorphic to the classical algebra (cf.\cite{Rovellir}). A comprehensive reference
is \cite{Thiemannr}.
\par
Some states which satisfy the Astekhar constraints are given by the
 loop representation, which can be introduced from the construct (depending both on
the gauge field $A$ and on a parametrized loop $\gamma$)
\be
 W (\gamma , A)\equiv tr\, P e^{\oint_{\gamma}A}
\ee
and a functional transform mapping functionals of the gauge field $\psi(A)$ into functionals
of loops, $\psi(\gamma)$:
\be
 \psi(\gamma)\equiv \int {\cal D}A\, W(\gamma,A)\psi(A)
\ee
When one divides by diffeomorphisms, it is found that
functions of knot classes (diffeomorphisms classes of smooth, non self-intersecting loops)
 satisfy all the constraints. 
 \par
Some particular states sought to reproduce smooth spaces at coarse graining are
the {\em Weaves}. It is not clear to what extent they also approach the conjugate variables (
that is, the extrinsic curvature) as well.
\par
In the presence of a cosmological constant the hamiltonian constraint reads:
\be
\epsilon_{ijk}E^{ai}E^{bj}(F^k_{ab}+\frac{\lambda}{3}\epsilon_{abc}E^{ck})=0
\ee
A particular class of solutions of the constraint \cite{Smolinc} are self-dual solutions of 
the form
\be
F^i_{ab}=-\frac{\lambda}{3}\epsilon_{abc}E^{ci}
\ee
Kodama (\cite{Kodama}) has shown that the Chern-Simons state 
 \be
 \psi_{CS}(A)\equiv e^{\frac{3}{2\lambda} S_{CS}(A)}
\ee 
is a solution of the hamiltonian constraint. He even suggested that the {\em sign} of the coarse
grained, classical  cosmological constant was always positive, irrespectively 
of the sign of the 
quantum 
parameter $\lambda$, but it is not clear whether this result is general enough.
 There is some concern \cite{Wittenk} 
that this state as such is not normalizable with the usual norm. It 
has been argued that
this is only natural, because the physical relevant norm must be very different from the 
na\"{\i}ve one
(cf. \cite{Smolin}) and indeed normalizability of the Kodama state has been suggested as a 
criterion
for the correctness of the physical scalar product(cf. for example the discussion in 
\cite{Smo}) or else that a euclidean interpretation could be given to it.
\par
Loop states in general (suitable symmetrized) can be represented as 
spin network (\cite{RS}) states: colored lines (carrying some $SU(2)$ representation) 
meeting at nodes where intertwining $SU(2)$ operators act. A beautiful graphical
representation of the group theory has been succesfully adapted for this purpose.
There is a clear relationship between this representation and the Turaev-Viro \cite{Turaev} 
invariants.
Many of these ideas have been foresighted by Penrose (cf. \cite{penrose}).
\par
There is also a 
path integral representation, known as  {\em spin foam} (cf.\cite{Baez}), a topological theory 
of colored surfaces representing the evolution of a spin network.
These are closely related to topological BF theories, and many independent generalizations 
have been proposed.
Spin foams can also be considered as an independent approach to the quantization of the 
gravitational 
field.(\cite{Barrett})
\par
In addition to its specific problems, this approach shares with all canonical approaches
to covariant systems
the problem of time. It is not clear its definition, at least in the absence of matter.
Dynamics remains somewhat mysterious; the hamiltonian constraint does not say in what sense
(with respect to what) the three-dimensional dynamics evolve.

\subsection{Big results of this approach.}
One of the main successes of the loop approach is that the
area (as well as the volume) operator is discrete. This allows, assuming that 
a black hole has been formed (which is a process that no one knows how to represent
in this setting), to explain the formula for the black hole entropy . The result
is expressed in terms of the Barbero-Immirzi parameter (\cite{RSS}). The physical meaning
 of this dependence is not well understood.
\par
 
It has been pointed out \cite{Bekenstein} that there  is a potential drawback
in all theories in which the area (or mass) spectrum is discrete with eigenvalues $A_n$ 
if the level spacing between eigenvalues $\d A_n$ is uniform because of the predicted thermal 
character
of Hawking's radiation. The explicit computations in the present setting, however, lead to
an space between (dimensionless) eigenvalues
\be
 \delta\, A_n\sim e^{-\sqrt{A_n}}, 
 \ee 
which seemingly avoids this set of problems.
  \par 
It has also been pointed out that \cite{Freidel} not only the spin foam, but almost
all other theories of gravity can be expressed as topological BF theories with constraints.
While this is undoubtely an intesting and potentially useful remark, it is
important to remember that the difference between the linear sigma model (a free field theory)
and the nonlinear sigma models is just a matter of constraints. This is enough to 
produce a mass gap
and asymptotic freedom in appropiate circumstances.


\section{Euclidean quantum gravity}
It can be boldly asserted that just by analogy with ordinary quantum field theory, the
wave functional of quantum gravity must be given by:
\be
\psi[h]\equiv \int_{g(\pd M)=h} {\cal D}g e^{-S_E[g]}
\ee
where we integrate over all riemannian metrics that obey the relevant boundary conditions,
and the Einstein-Hilbert action has to be supplemented with boundary terms. This approach
is problematic from the very beginning, due to the fact that the Wick
analytic continuation of a lorentzian space-time is not riemannian in general
(not even real), so that the whole setup seems to demand the study of real sections
in a  complex formulation. The point of view put forward by Hawking and collaborators 
\cite{Hawking euc} is that the needed analytical continuations could be hopefully made
 after Green's functions are evaluated.
\par
There however is a well-known mathematical theorem of Markov (explained for
physicists  in \cite{Alvarez:1991rw})
 asserting that there is no algorithmic way of predicting when two arbitrary
manifolds are homeomorphic :$M_1\sim M_2$.
The problem stems  essentially from the fundamental group: any finitely presented group 
can be the $\pi_1 (M)$ of a four-dimensional manifold, $M$.
So one proof of the result is to simply write down a family of groups $G_k$ such that we cannot
algorithmically recognise when $G=\{e\}$.
There are, in addition, further subtleties with the diffeomorphism 
class in $d=4$: there is a uncountable
set of non equivalent differentiable structures in $\mathbb{R}^4$: the so-called {\em exotic}
$\mathbb{R}^4$ (cf. \cite{Brans} for a physical approach; a relevant recent reference is 
\cite{Pfeiffer}) .
\par

 Working in lorentzian signature, a Hamiltonian path integral can be dreamt of, where a
 functional integral is performed over three-dimensional geometries (cf. \cite{Alvarez:1992mb} 
only. Here the situation
is slightly better: it seems that there is recent progress in the proof of 
Thurston's geometrization conjecture, which implies in particular Poincar\'e's conjecture,
and which explains all three-dimensional manifolds in terms of eight different {\em geometries}.
Incidentally, the work of Perelman (\cite{Perelman}) uses what mathematicians call the 
Ricci flow,
which is {\em exactly} the flow of the renormalization group of the sigma model associated
to the bosonic string in a curved background.
\par
Let us finally comment that even if the basic theory of Nature is topological
one needs to {\em enumerate} topologies to discriminate between different ones.
Besides, topological symmetry has to be broken al low energies.

\par
In order to reach a probabilistic interpretation, a scalar product ought to be defined.
The one which is naturally associated to the Wheeler-de Witt equation is not
positive-definite, so this remains as an open problem in this approach.
\par
Were somebody apply these ideas to the whole Universe (the so called Quantum Cosmology)
there are other problems in store.
It is not clear what is the physical interpretation of probabilities associated to a single
 event. A related problem is the one of the physical interpretation of Quantum 
Mechanics  without classical observers. Many people have related this to the decoherence
mechanisms (cf. for example \cite{Hartle}) but it seems to me that the situation is still to
be clarified.

\section{ Perturbative (graviton) approach}

A much more modest approch is to study gravitons as ordinary (massless, spin two) particles
in Minkowski space-time.
\be
g_{\a\b}=\bar{g}_{\a\b}+\kappa h_{\a\b}
\ee
It seems to many people (including the author) that this is at least a preliminary step before
embarking in more complicated adventures. As a quantum field theory, quantum general relativity
has got a dimensionful coupling : $d(\kappa)=-1$, which means that it is not renormalizable in 
the usual sense of the word.
\par
In spite of this, the theory is 
one loop finite on shell , as was shown in a brilliant calculation by G. 't Hooft and M. Veltman
(\cite{hooft}). They computed the counterterm:
\be
\Delta L^{(1)} = \frac{\sqrt{\bar{g}}}{\e}\frac{203}{80}\bar{R}^2
\ee
\par
No more miracles are expected for higher loops, and none happen. 
Goroff and Sagnotti (\cite{Goroff}) were the first to show that
to two loops,
\be
\Delta L^{(2)}=\frac{209}{2880 (4\pi)^4}
\frac{1}{\e}\bar{R}^{\a\b}\,_{\g\d}\bar{R}^{\g\d}\,_{\rho\sigma}
\bar{R}^{\rho\sigma}\,_{\a\b}
\ee

The general structure of perturbation theory is governed by the fact we have just mentioned
that the coupling constant
is dimensionful.
A general diagram  will then behave in the s-channel as  $\kappa^n s^n$ and 
counterterms  as:
\be
\Delta L\sim \sum \int \kappa^n R^{(2+ n/2)}
\ee
(where a symbolic notation has been used, packing all invariants with the same dimension;
for example, $R^2$ stands for an arbitary combination of $R^2$, $R_{\a\b}R^{\a\b}$ and
$R_{\a\b\g\d}R^{\a\b\g\d}$)
conveying the fact that  that the theory is non-renormalizable.
\par
It may however be pondered whether
effective lagrangians are really 
useful for $E<< m_P$. This possibility has been forcefully explored by
 Donoghue and collaborators (cf. \cite{Donoghue}).
There are some caveats: for example, when horizons are present, it seems necessary
in order to be able to apply these ideas, to use some particular foliations, the
so called {\em nice slices}). The mere fact that we are unable to predict
 the cosmological constant
(which is the mother of all infrared problems) means that our understanding has ample room
for improvement.
\par
Could it be that in spite of the fact that general relativity is not renormalizable,
there is a non perturbative sector in which the theory makes sense as a quantum theory?
First of all, were that true, it would be most remarkable: there are no known QFT which are
 defined only noperturbatively. Besides, at the classical level, perturbation theory
works wonderfully, and there is indeed a whole framework, the parametrized post-newtonian
(PPN) formalism to discriminate netween alternate theories of gravity. It is then most unclear 
why at the quantum (and only there) level perturbation theory should fail.

\par
We want to mention in closing this chapter, some fascinating relationships uncovered by
Z. Berm and collaborators (cf. \cite{Bern}) between purely field-theoretical 
S matrix elements in (super)gravity 
and gauge theories: the so called {\em Gravity=Gauge $\times$ Gauge} conjecture.
In spite of several attempts, it is not clear how this can be understood from the
 Einstein-Hilbert action.
The relationship is of course automatic in strings, because closed string amplitudes (which
inclñude the graviton) are
given by products of open string amplitudes (which contain the gauge fields).The KLT relations
(\cite{Kawai}) are a quantitative formulation of this fact.

\par
 In view of all this, one can try to study particular extensions of the Einstein-Hilbert
actions. Modifications quadratic in the curvature improve renormalizability (\cite{Julve}), 
but have problems with unitarity at a very fundamental level (\cite{Stelle}).
\par
Local supersymmetry is expected to improve the ultraviolet behavior through cancellations
between fermionic and bosonic degrees of freedom. In spite of that, some infinities
are allowed by the symmetries of the problem, and are thus expected to appear; for example
in extended supergravities this is expected to happen at loop order $L>\frac{10}{D-2}$ 
in the maximally supersymmetric case in which there are  32 supercharges.
\par
The (sad) conclusion of all this is that ordinary QFT (with a finite number of fields) 
does not work, even for describing small (quantum) ripples in Minkowski space.

\section{Strings}
It should be clear by now that we probably still do not know
what is exactly the problem to which string theories are the answer. At any rate, the starting 
point is that all elementary particles are viewed as quantized excitations of a one 
dimensional object,
the string, which can be either open (free ends) or closed (a loop). Excellent books are 
avaliable, such as \cite{Greens}\cite{Polchinskis}.
\par
String theories enjoy a convoluted history. Their origin can be traced to the
Veneziano model  of strong interactions. A crucial step was the reinterpretation 
by Scherk and Schwarz 
(\cite{Scherk}) of the massless spin two state in the closed sector (previously 
thought to be related to the 
Pomeron) as the  
graviton and consequently of the whole string theory as
a potential 
theory of quantum gravity, and potential unified theories of all interactions. Now the wheel has
made a complete turn, and we are perhaps back
through the Maldacena conjecture (\cite{Maldacena}) to a closer relationship than previously 
thought 
with ordinary gauge theories.
\par

\begin{figure}[!ht] 
\begin{center} 
\leavevmode 
\epsfxsize= 12cm

\epsffile{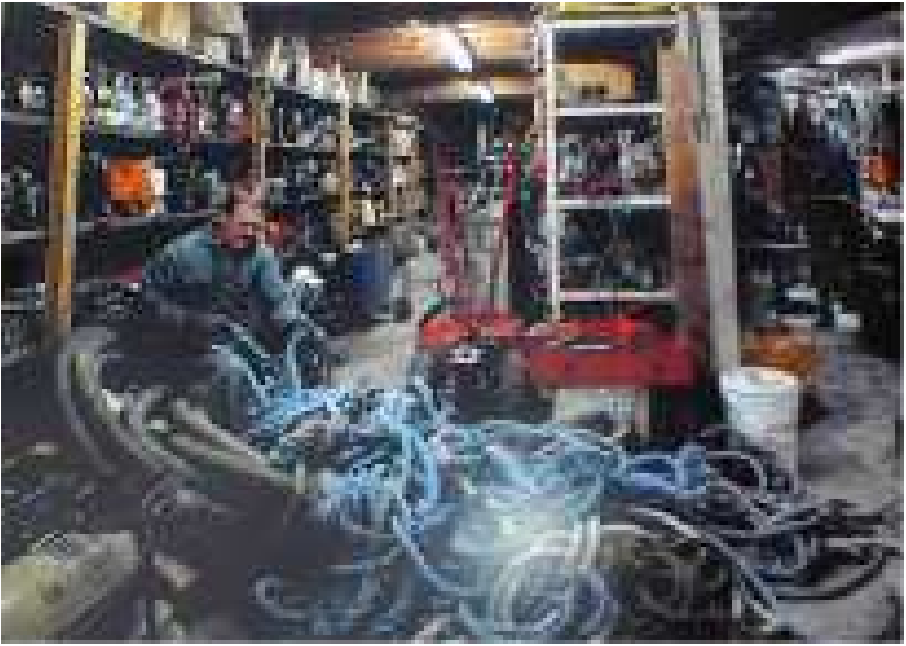} 
\caption{String theorist at work.}
\label{fig:3}   
\end{center} 
\end{figure}

From a certain point of view, their dymamics is determined by a two-dimensional non-linear sigma
model, which geometrically is a theory of imbeddings of a two-dimensional surface (the world
sheet of the string) to a (usually ten-dimensional) target space:
\be
x^{\m}(\xi): \Sigma_2\rightarrow M_n
\ee
There are two types of interactions to consider.
Sigma model interactions (in a given two-dimensional surface) 
are defined as an expansion in powers of momentum, where a new
dimensionful parameter, $\a^{\prime}\equiv l_s^2$ sets the scale. This scale
is {\em a priori} believed to be of the order of the Planck length. The first terms in 
the action
always include a coupling to the massless backgrounds: the spacetime metric, the two-index 
Maxwell
like field known as the Kalb-Ramond or $b$-field, and the dilaton. To be specific,
\be
 S= \frac{1}{l_s^2}\int_{\Sigma_2}g_{\mu\nu}(x(\xi))\partial_a x^{\mu}(\xi)\partial_b 
x^{\nu}(\xi)\gamma^{ab}(\xi)
+\ldots 
\ee 
There are also string interactions, (changing the two-dimensional surface) 
proportional to the string coupling constant, $g_s$, whose
variations are related to the logarithmic variations of the dilaton field.
Open strings (which have gluons in their spectrum) {\em always} contain closed strings 
(which have gravitons in their spectrum) as intermediate
states in higher string order ($g_s$)  corrections. This interplay open/closed is one of the 
most fascinating aspects of the whole string theory.
\par
It has been discovered by Friedan (cf. \cite{Friedan}) that in order for the quantum theory
to be consistent with all classical symmetries (diffeomorphisms and conformal invariance), the 
beta function of
the generalized couplings \footnote{There are corrections coming from both dilaton and 
Kalb-Ramond
fields. The quoted result is the first term in an expansion in derivatives, with 
expansion parameter
 $\a^{\prime}\equiv l_s^2$.} must vanish:
\be
 \beta (g_{\mu\nu})=R_{\mu\nu}=0 
\ee 
  This result remains until now as one of the most important ones in string theory, hinting 
at a 
deep relationship between Einstein's equations and the renormalization group.
   
Polyakov (\cite{Polyakov}) introduced the so called {\em non-critical strings} which have
in general a two-dimensional cosmological constant (forbidden otherwise by Weyl invariance).
The dynamics of the conformal mode (often called Liouville in this context) is, however, 
poorly understood.
     
\par   
 
Fundamental strings live in  D=10 spacetime dimensions, and so a Kaluza-Klein mecanism of sorts
must be at work in order to explain why we only see four non-compact dimensions at low energies.
Strings have in general tachyons in their spectrum, and the only way to construct seemingly
consistent string theories (cf. \cite{Gliozzi}) 
is to project out those states, which leads to supersymmetry. This means in turn that all 
low energy
predictions heavily depend on the supersymmetry breaking mechanisms. 
\par
  
 String perturbation theory is probably well defined although a full proof is not available.
\par
Several stringy symmetries are believed to be exact: T-duality, relating large and small 
compactification
volumes, and $S$-duality,
relating the strong coupling regime with  the weak coupling one.
Besides, extended configurations ({\em D branes}); topological defects in which open strings can
end are known to be important \cite{Polchinski}. They couple to Maxwell-like fields which 
are p-forms
called Ramond-Ramond (RR) fields.
These dualities \cite{Hull} relate  all five string theories (namely, Heterotic $E(8)\times 
E(8)$, 
Heterotic $SO(32)$, Tipe $I$, $IIA$ and $IIB$) and it is conjectured that there is an 
unified eleven -dimensional theory, dubbed $M$-theory
of which ${\cal N}=1$ supergravity in $d=11$ dimensions is the low energy limit.

\subsection{Big results}
Perhaps the main result is that graviton physics in flat space is well defined for the first 
time, and
this is no minor accomplishment.
\par
Besides, there is evidence that at least some geometric singularities are harmless
in the sense that strings do not feel them.
Topology change amplitudes do not vanish in string theory.
\par
The other Big Result \cite{Strominger} is that one can
correctly count states
 of extremal black holes
 as a function of charges. This is at the same time astonishing and disappointing.
It clearly depends strongly on the objets being BPS states (that is, on supersymmetry),
and the result has not been extended to non-supersymmetric configurations.
On the other hand, as we have said, it {\em exactly} reproduces the entropy as a 
function of a sometimes large number of charges, without any adjustable parameter.
     
\subsection{The Maldacena conjecture}
 Maldacena \cite{Maldacena} proposed as a conjecture that $IIB$ string theories in a background
$AdS_5\times S_5$ with  common radius $l\sim l_s (g_s N)^{1/4}$ and N units of RR flux 
that is,  $\int_{S_5} F_5 =N$ (which implies that $F_5\sim \frac{N}{r^5}$)
is equivalent to a four dimensional ordinary gauge theory in flat four-dimensional Minkowski
space, namely ${\cal N}=4$ super Yang-Mills with gauge group $SU(N)$ and coupling constant
$g=g_s^{1/2}$.
 \par

Although there is much supersymmetry in the problem and the kinematics largely determine 
correlators,
(in particular, the symmetry group $SO(2,4)\times SO(6)$ is realized as an isometry group on the
gravity side and as an $R$-symmetry group as well as conformal invariance on the gauge 
theory side)
this is not fully so \footnote{The only correlators that are completely determined
through symmetry are the two and three-point functions.}
and the conjecture has passed many tests in the semiclassical approximation
to string theory.

The action of the RR field, given schematically by $\int F_5^2$, scales as $N^2$, whereas the 
ten-dimensional 
Einstein-Hilbert $\int R$, depends on the overall geometric scale as the eighth power of the
common radius, $l^8$. The 't Hooft coupling is
 $\lambda = g^2 N \sim \frac{l^4}{l_s^4}$  and the tenth dimensional Newton's constant
is  $\kappa_{10}^2\sim G_{10}\sim l_p^8=g_s^2 l_s^8\sim\frac{l^8}{N^2}$.
\par  
 If we consider the effective five dimensional theory after compactifying 
on a five sphere of radius $r$, the RR term yields a negative contribution $\sim -
(\frac{N}{r^5})^2 r^5$ ,whereas the positive curvature of the five sphere $S^5$ gives a positive
contribution, $\sim \frac{1}{r^2} r^5$. The competition between these two terms in the effective
potential is responsible for the minimum with negative cosmological constant.

\par
The way the dictionary works  in detail \cite{Witten} is that the supergravity action 
corresponding to fields 
with prescribed
boundary values is related to gauge theory correlators of certain gauge invariant operators 
corresponding to the particular field studied:
\be
e^{- S_{sugra}[\Phi_i]}\bigg|_{\Phi_i|_{\partial AdS}=\phi_i}= 
<e^{\int {\cal O}_i\phi_i}>_{CFT}
\ee

\par
 This is the first time that a precise holographic description of spacetime in terms of a 
(boundary) 
gauge theory is proposed and, as such it is of enormous potential interest. 
It has been conjectured 
by 't Hooft \cite{'tHooft} and further developed by Susskind \cite{Susskind} that there 
should be
much fewer degrees of freedom in quantum gravity than previously thought. The conjecture 
claims that
it should be enough with one degree of freedom per unit Planck surface in the two-dimensional
 boundary
of the three-dimensional volume under study. The reason for that stems from an analysis of the 
Bekenstein-Hawking \cite{Bekenstein}\cite{Hawking} 
entropy associated to a black hole, given in terms of the
two-dimensional area $A$ \footnote{The area of the horizon for a Schwarzschild black hole 
is given by:
\be
A=\frac{8\pi G^2}{c^4}M^2
\ee
}
of the horizon by
\be
S=\frac{ c^3}{4 G\hbar}A.
\ee

This is  a deep result indeed, still not fully understood.
\par
It is true on the other hand that the Maldacena conjecture has only been checked for the 
time being 
in some corners of 
parameter space, namely when strings can be approximated by supergravity in the appropiate 
background.

\begin{figure}[!ht] 
\begin{center} 
\leavevmode 
\epsfxsize= 10cm

\epsffile{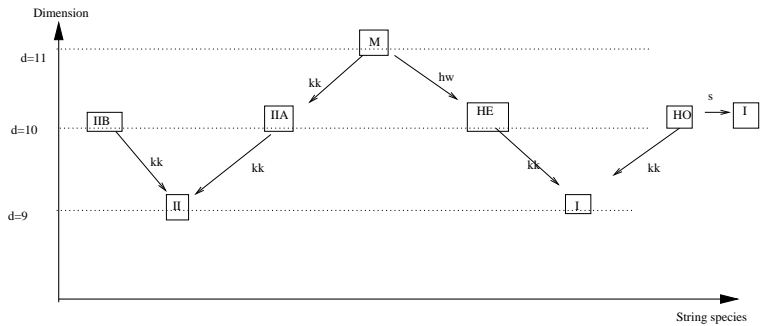} 
\caption{Conjectured relationships between string theories in different dimensions.}
\label{fig:4}   
\end{center} 
\end{figure}

%

%

\section{Dualities and branes}
The so- called {\em T-duality} is the simplest of all dualities and the 
only one which can be shown
to be true, at least in some contexts. At the same time it is a very stringy
characteristic, and depends in an essential way on strings being extended
objects.  In a sense, the web of dualities
rests on this foundation, so that it is important to understand 
clearly the basic 
physics involved. Let us consider strings living on an external space with one 
compact dimension, which we shall call $y$, with topology $S^1$ and radius $R$. 
The corresponding field in the imbedding of the
string, which we shall call $y$
({\it i.e.} we are dividing the target-spacetime dimensions as $(x^{\m},y)$,
where $y$ parametrizes the circle), has then the
possibility of winding around it:
\be
y(\sigma + 2\pi,\tau) = y(\sigma, \tau) + 2\pi R m \; .
\ee
A closed string can close in general up to an isometry of
the external spacetime.
\par
The zero mode expansion of this coordinate (that is, forgetting about
oscillators) would then be
\be
y = y_{c} + 2 p_{c} \tau + m R \sigma \; .
\ee
Canonical quantization leads to $[y_{c},p_{c}] = i$, and single-valuedness
of the plane wave $e^{iy_{c}p_{c}}$ enforces as usual $p_{c}\in \mathbb{Z}/R$,
so that $p_{c} = \frac{n}{R}$.
\par
The zero mode expansion can then be organized into left and right movers in the
following way
\bea
y_L (\tau +\sigma)&=& y_c/2 + 
     \left(\frac{n}{R} + \frac{mR}{2}\right) (\tau + \sigma) \; ,\nonumber\\
y_R (\tau -\sigma)&=& y_c/2 + 
    \left(\frac{n}{R} - \frac{mR}{2}\right) (\tau - \sigma) \; .
\eea
The mass shell conditions reduce to
\bea
m_L^2 &=& \frac{1}{2}\left(\frac{n}{R} + 
          \frac{mR}{2}\right)^2 + N_L -1 \; ,\nonumber\\
m_R^2 &=& \frac{1}{2}\left(\frac{n}{R} - \frac{mR}{2}\right)^2 + N_R -1 \; .
\eea
Level matching, $m_L = m_R$, implies that there is a relationship between
momentum and winding numbers on the one hand, and the oscillator 
excess on the other
\be
N_R - N_L \;=\;  nm \; .
\ee
At this point it is already evident that the mass formula is invariant
under
\be
R\; \rightarrow\;  R^{*}\equiv 2/R \; ,
\ee
and exchanging momentum and winding numbers. This is the simplest instance of 
{\em T-Duality}.
\par
On the other hand, it is an old observation (which apparently originated in 
Schr\"odinger) that Maxwell's 
equations are almost symmetrical with respect to
interchange between electric and magnetic degrees of freedom. This idea was explored by Dirac
and eventually lead to the discovery of the consistency conditions that have to be fulfilled
if there are magnetic monopoles in nature. The fact that nonsingular magnetic monopoles 
appear as classical solutions in some gauge theories led further support to this duality
viewpoint. In order to be able to make a consisting conjecture, first put forward by Montonen
and Olive \cite{Montonen}, supersymmetry is needed, as first remarked by Osborn \cite{Osborn}. 
\par
Now in strings there are the so- called Ramond-Ramond (RR) fields, which are p- forms
of different degrees. In the same way that one forms (i.e., the Maxwell field) couples to 
charged particles  that is, from the spacetime point of view, to objects of dimension 0
with one-dimensional trajectories, a p-form
\be
A_{\m_1\ldots\m_p}
\ee
would couple to a $(p-1)$-dimensional object, whose world history is described by a 
$p$-dimensional
hypersurface
\be
x^{\m}= x^{\m}(\xi_1\ldots \xi_p)
\ee
These objects are traditionally denoted by the name $p$-branes (it all originated in a 
dubious joke). That is, ordinary particles are $0$-branes, a string is a $1$-brane, a membrane
is a $2$-brane, and so on.
\par
Dualities relate branes of different dimensions in different theories; this means that
if one is to take this symmetry seriously, it is not clear at all that strings are the
more fundamental objects: in the so called $M$-theory branes appear as fundamental as strings.
\par
If we are willing to make the hypothesis
that supersymmetry is not going to
be broken whilst increasing the coupling constant, $g_s$,
some astonishing conlusions
can be drawn. Assuming this, massless quanta can become massive
as $g_s$ grows only if their number, charges and spins are such that
they can combine into massive multiplets  
(which are all larger than the irreducible massless ones). 
The only remaining issue, then, is whether
any other massless quanta can appear
at strong coupling.
\par
Now, in the IIA string theory there are states associated to the Ramond-Ramond (RR)
one form, $A_1$, namely the D-0-branes, whose tension goes as 
$m\sim\frac{1}{g_s}$. This clearly gives new massless states in the strong 
coupling limit.
\par
There are reasons\footnote{In particular: The fact
that there is the possibility of a central extension in the IIA algebra,
related to the Kaluza-Klein compactification of the d=11 Supergravity
algebra.} to think that this new massless states  are the first
level of a Kaluza-Klein tower associated to compactification on a circle
of an 11-dimensional theory.
Actually, assuming an 11-dimensional spacetime with an isometry 
$k=\frac{\partial}{\partial y}$, an Ansatz which exactly reproduces the 
dilaton factors of the IIA string is
\be
ds^{2}_{(11)} = e^{\frac{4}{3}\phi}(dy - A^{(1)}_{\m} dx^{\m})^2 + 
e^{-\frac{2}{3}\phi}g_{\m\n}dx^{\m}dx^{\n} \; .
\ee 
Equating the two expressions for the D0 mass,
\be
\frac{1}{g_s} = \frac{1}{R_{11}} \; ,   
\ee
leads to $R_{11} = e^{\frac{2}{3}\phi} = g_A^{2/3}$.
\par
This means that a new dimension appears at strong coupling, and this dimension is 
related to the dilaton. The only reason why we do not see it at low energiew is 
precisely because of the smallness of the string coupling, related directly to the
dilaton field.
The other side of this is that this eleven dimensional theory, dubbed {\em M-theory}
does not have any weak coupling limit; it is always strongly coupled. Consequently, not much
is known on this theory, except for the fact that its field theory, low curvature limit
is ${\cal N}=1$ supergravity in $d=11$ dimensions.
\par
All supermultiplets of massive one-particle states of the IIB string
supersymmetry algebra contain states of at least spin 4. 
This means that under the previous set of hypothesis, the set of 
massless states at weak coupling must
be exactly the same as the corresponding set at strong coupling.
This means that there must be a symmetry mapping weak coupling into 
strong coupling.
\par
There is a well-known candidate for this symmetry:
Let us call, as usual, $l$ the RR scalar and $\phi$ the dilaton (NSNS).
We can pack them together into complex scalar
\be
S \;=\;  l \,+\, i e^{-\frac{\phi}{2}} \; .
\ee
The IIB supergravity action in d=10 is invariant under the $SL(2,\mathbb{R})$
transformations
\be
S \rightarrow \frac{a S + b}{ c S + d} \; ,
\ee
if at the same time the two two-forms, $B_{\m\n}$ (the usual, ever-present,
NS field), and $A^{(2)}$, the RR field transform as
\bea
\left( \begin{array}{c}
B\\
A^{(2)}
\end{array}\right ) \rightarrow 
\left( \begin{array}{cc}
d & - c \\
- b & a\end{array}\right )
\left ( \begin{array}{c}
B\\
A^{(2)}
\end{array}\right) \; , 
\eea
Both the, Einstein frame, metric $g_{\m\n}$
and the four-form $A^{(4)}$ are inert under this
$SL(2,\mathbb{R})$ transformation.
\par
A discrete subgroup $SL(2,\mathbb{Z})$ of the full classical $SL(2,\mathbb{R})$
is believed to be an exact symmetry of the full string theory. 
The exact imbedding of the discrete subgroup in the full $SL(2,\mathbb{R})$ 
depends on the
vacuum expectation value of the RR scalar.
\par
The particular transformation
\bea
g = \left(
\begin{array}{cc}
 0& 1\\
- 1& 0\end{array}
\right) \; ,
\eea
maps $\phi$ into $ - \phi$ (when $l =0$),
and $B$ into $A^{(2)}$. 
This means that the string coupling
\be
g_s\rightarrow \frac{1}{g_s}
\ee
This is a strong/weak coupling type of duality, similar to the electromagnetic duality
in that sense .The standard name for it is  an 
{\em S-duality} type of transformation, mapping the ordinary string with NS charge,
to another string with RR charge (which then must be a D-1-brane,
and is correspondingly called a {\em D-string}), and, from there, 
is connected to all other D-branes by T-duality.
\par

Using the fact that upon compactification on $S^1$, IIA at $R_A$ is
equivalent to IIB at $R_B\equiv 1/R_A$, and the fact that the effective action
carries a factor of $e^{-2\phi}$ we get
\be
R_A g_B^2 = R_B g_A^2 \; ,
\ee
which combined with our previous result, $g_A = R_{11}^{3/2}$ implies 
that $ g_B = \frac{R_{11}^{3/2}}{R_A}$.
Now the Kaluza-Klein Ansatz implies that from the eleven dimensional 
viewpoint the compactification radius is measured as
\be
R_{10}^2 \equiv R_A^2 e^{- 2\phi/3} \; ,
\ee
yielding
\be
g_B = \frac{R_{11}}{R_{10}} \; .
\ee
\par
 From the effective actions written above it is easy to check 
that there is a (S-duality type) field transformation mapping
the SO(32) Type I open string into the SO(32)
Heterotic one namely
\bea
g_{\m\n}&&\rightarrow e^{-\phi} g_{\m\n}^{Het} \; ,\nonumber\\
\phi&&\rightarrow - \phi \; ,\nonumber\\
B'&&\rightarrow B \; .
\eea
This means that physically there is a strong/weak coupling duality, because 
coupling constants of the compactified theories  would
be related by
\bea
g_{het}&& = 1/g_I \; ,\nonumber\\
R_{het}&& = R_I/g_I^{1/2} \; .
\eea

\section{Summary: the state of the art in quantum gravity}
In the loop approach one is working
 with  nice candidates for a quantum
  theory. The theories are interesting, probably related to topological field theories 
(\cite{Blau})
and background independence as well as diffeomorphism invariance are clearly implemented.
On the other hand, it is not clear that their low energy limit  is related to 
Einstein gravity.
    
\par 
 Strings start from a perturbative approach more familiar to a particle physicist.
However, they carry all the burden of supersymmetry and Kaluza-Klein. It has proved to be very
difficult to study nontrivial non-supersymmetric dynamics.
\par
Finally, and this applies to  all approaches, the holographic ideas seem intriguing; there are
many indications of a deep relationship between gravity and gauge theories.
\par
We would like to conclude by insisting on the fact that although there is not much we know for 
sure on quantum effects on the gravitational field, even
the few things we know are a big feat, given the difficulty to do physics 
without experiments.

\par

 Progress could be made if we could derive semiclassical gravity in such a way that 
corrections to it can be reliably estimated, for example
\be
 R_{\m\n}-\frac{1}{2}R g_{\m\n}=2\kappa^2 <\psi|\hat{T}_{\m\n}|\psi>+ \frac{1}{L^2}\Delta,
\ee
when working at a certain scale of distances, say $L$.
In order to understand those equations, we would had to know something about the operator
of which the first member is the expectation value; something about the state on which
the expectation value is computed (In particular, if it is the vacuum, how is it to be defined?)
and finally, something about the definition of the energy momentum tensor as a 
composite operator.    
A question of obvious physical interest is the estimate of the size of the corrections:
Is the expected error at a given scale of distance $L$
\be
\Delta\sim \frac{\hbar G}{c^3 L^2}
\ee
or, does it depend of the characteristic energy of the source?
\be
\Delta\sim \frac{G E^2}{\hbar c^5}
\ee 
measured with respect to what?
\par
It is painfully clear that there is still a large margin for 
improving our understanding of effective quantum field theories. For example, there is 
still no convincing derivation of  Hawking radiation without transplanckian modes appearing
at some point (this particular example is related to the existence of the nice 
slices mentioned above). Besides, we do not understand the cosmological constant, 
which is clearly related to the estimate of $\Delta$.
\par

The observational prospects are rather poor. In many models, in particular in the loop 
approach ( and also in strings, with some qualifications) deviations from the lorentzian
dispersion relations are expected:
\be 
E^2 = \vec{p}^2 + m^2 + E^2\sum_{n=1} c_n (\frac{E}{m_P})^n
\ee
Other contributions will undoubtly analyze those in much more detail. Let us now simply
mention that
noncommutative models make similar predictions.
\par
 Winding states are stringy phenomena, and its observation would be very interesting.
Stringy predictions, however, are in general difficuly to disentangle from  predictions
of supersymmetry (SUSY). Namely, SUSY has to be broken, and this scale spoils almost 
all differences between strings and QFT models.
\par   

With the great triumph of particle physics at the end of the seventies, namely 
the {\em experimental} discovery of the intermediate bosons related to electroweak interactions,
the standard model was confirmed in all its essential traits, waiting only for the 
Higgs to be discovered (at LHC?) and the theoretical effort has concentrated in more and 
more speculative topics, and experimental 
guidance has become correspondingly scarce. The net result is that, even more so that in the old
days of the hunting for the theory of strong interactions, theoretical physics is divided in
almost disconnected clans. 

All this is even more true when talking about quantum gravity,
a paradise of speculation. 

This is the reason why all efforts such as the one in the present
workshop, aiming at making contact with experiment and/or observation 
are welcome, and will eventually
redirect physics on a healthier track when we  learn to recognise the physically 
relevant facts that  presumably lie in front of our eyes.

\section{Acknowledgements}
I am grateful to Giovanni Amelino-Camelia and to Jerzy Kowalski-Glikman for organizing 
such a stimulating
workshop and for allowing me to take part in it. I am also 
grateful to Carlos Mu\~noz for lending his picture of a stringer .
Finally, I am indebted to Jorge Conde for reading the manuscript.

\appendix



\begin{thebibliography}{99}

\bibitem{ADM}R. Arnowitt, S. Deser, and C. W. Misner,
`` Canonical Variables for General Relativity''
Phys. Rev. 117, 1595-1602 (1960)  


\bibitem{Alfaro}
J.~Alfaro and G.~Palma,
``Loop quantum gravity corrections and cosmic rays decays,''
Phys.\ Rev.\ D { 65} (2002) 103516
[arXiv:hep-th/0111176].
\\
``Loop quantum gravity and ultra high energy cosmic rays,''
Phys.\ Rev.\ D { 67} (2003) 083003
[arXiv:hep-th/0208193].



\bibitem{Alvarez}
E.~Alvarez,
``Quantum Gravity: A Pedagogical Introduction To Some Recent Results,''
Rev.\ Mod.\ Phys.\  { 61} (1989) 561.

\bibitem{Alvarez:gd}
E.~Alvarez,
``Low-Energy Effects Of Quantum Gravity,''
FTUAM-89-15
{\it Lectures given at Mtg. on Recent Developments in Gravitation, Barcelona, Spain, 
Sep 5-8, 1989}

\bibitem{Alvarez:1991rw}
E.~Alvarez,
``Some general problems in quantum gravity,''
CERN-TH-6257-91
{\it Lectures at the 22nd Gift Int. Seminar in Theoretical Physics: Quantum Gravity and Cosmology, S. Feliu De Guixols, Jun 3-6, 1991}

\bibitem{Alvarez:1992mb}
E.~Alvarez,
``Some general problems in quantum gravity. 2. The Three-dimensional case,''
Int.\ J.\ Mod.\ Phys.\ D { 2} (1993) 1
[arXiv:hep-th/9211050].






\bibitem{Alvarezz}
E.~Alvarez, L.~Alvarez-Gaume and Y.~Lozano,
``An introduction to T duality in string theory,''
Nucl.\ Phys.\ Proc.\ Suppl.\  { 41} (1995) 1
[arXiv:hep-th/9410237].


\bibitem{Alvarez-Gaume}
L.~Alvarez-Gaume and M.~A.~Vazquez-Mozo,
``Topics in string theory and quantum gravity,''
arXiv:hep-th/9212006.


\bibitem{Antoniadis}
I.~Antoniadis, N.~Arkani-Hamed, S.~Dimopoulos and G.~R.~Dvali,
``New dimensions at a millimeter to a Fermi and superstrings at a TeV,''
Phys.\ Lett.\ B { 436} (1998) 257
[arXiv:hep-ph/9804398].


\bibitem{Ashtekar}
A.~Ashtekar,
``New Hamiltonian Formulation Of General Relativity,''
Phys.\ Rev.\ D { 36} (1987) 1587.



\bibitem{Astekhars}
A. Ashtekar, ``Quantum Geometry And Black Hole Entropy'',
 gr-qc/0005126 [=Adv.Theor.Math.Phys. 4 (2000) 1]\\
``Quantum Geometry Of Isolated Horizons And Black Hole Entropy'',
 Phys.Rev. D57 (1998) 1009 [=gr-qc/9705059]

\bibitem{Baez}
J.Baez, ``An Introduction To Spin Foam Models Of Quantum Gravity And Bf Theory'', gr-qc/0010050


\bibitem{Barbero}
J.~F.~Barbero,
``Real Ashtekar variables for Lorentzian signature space times,''
Phys.\ Rev.\ D { 51} (1995) 5507
[arXiv:gr-qc/9410014].


\bibitem{Barrett}
J.~W.~Barrett and L.~Crane,
``A Lorentzian signature model for quantum general relativity,''
Class.\ Quant.\ Grav.\  { 17} (2000) 3101
[arXiv:gr-qc/9904025].



\bibitem{Bekenstein}
J. Bekenstein, {\em Black  Holes And Entropy}
Phys.Rev. D7 (1973) 2333\\
(with V. Mukhanov), ``Spectroscopy Of The Quantum Black Hole'', Commun.Math.Phys. 125 (1989) 417

\bibitem{Bern}
Z.~Bern,
``Perturbative quantum gravity and its relation to gauge theory,''
Living Rev.\ Rel.\  {\bf 5} (2002) 5
[arXiv:gr-qc/0206071].



\bibitem{Blau}
M. Blau, ``Topological Gauge Theories Of Antisymmetric Tensor Fields'',
 hep-th/9901069 [=Adv.Theor.Math.Phys. 3 (1999) 1289]


\bibitem{Brans}
C.~H.~Brans,
``Exotic smoothness structures in physics,''
{\it Prepared for 1st Mexican School on Gravitation and Mathematical Physics, Guanajuato, Mexico, 12-16 Dec 1994}





\bibitem{Coleman}
S.~R.~Coleman and S.~L.~Glashow,
``High-energy tests of Lorentz invariance,''
Phys.\ Rev.\ D { 59} (1999) 116008
[arXiv:hep-ph/9812418].



\bibitem{Dewitt}
B.~S.~Dewitt,
``Quantum Theory Of Gravity. 1. The Canonical Theory,''
Phys.\ Rev.\  {\bf 160} (1967) 1113.\\
``Quantum Theory of Gravity. II. The Manifestly Covariant Theory'',
                          Phys. Rev. 162, 1195-1239 (1967).


\bibitem{Donoghue}
J.~F.~Donoghue,
``General Relativity As An Effective Field Theory: The Leading Quantum
Corrections,''
Phys.\ Rev.\ D {\bf 50} (1994) 3874
[arXiv:gr-qc/9405057].



\bibitem{Duff}
M.~J.~Duff,
``Inconsistency Of Quantum Field Theory In Curved Space-Time,''
ICTP/79-80/38
{\it Talk presented at 2nd Oxford Conf. on Quantum Gravity, Oxford, Eng., Apr 1980}




\bibitem{Freidel}
L.~Freidel, K.~Krasnov and R.~Puzio,
``BF description of higher-dimensional gravity theories,''
Adv.\ Theor.\ Math.\ Phys.\  { 3} (1999) 1289
[arXiv:hep-th/9901069].

\bibitem{Smo}
L.~Freidel and L.~Smolin,
``The linearization of the Kodama state,''
arXiv:hep-th/0310224.


\bibitem{Friedan}
D.~Friedan,
``Nonlinear Models In Two Epsilon Dimensions,''
Phys.\ Rev.\ Lett.\  { 45} (1980) 1057.





\bibitem{Gliozzi}
F.~Gliozzi, J.~Scherk and D.~I.~Olive,
``Supersymmetry, Supergravity Theories And The Dual Spinor Model,''
Nucl.\ Phys.\ B { 122} (1977) 253.


\bibitem{Goroff}
M.~H.~Goroff and A.~Sagnotti,
``The Ultraviolet Behavior Of Einstein Gravity,''
Nucl.\ Phys.\ B {\bf 266}, 709 (1986).




\bibitem{Green}
M.~B.~Green and J.~H.~Schwarz,
``Anomaly Cancellation In Supersymmetric D=10 Gauge Theory And Superstring Theory,''
Phys.\ Lett.\ B { 149} (1984) 117.

\bibitem{Greens}
M.~B.~Green, J.~H.~Schwarz and E.~Witten,
``Superstring Theory. Vol. 1: Introduction,''
\\
``Superstring Theory. Vol. 2: Loop Amplitudes, Anomalies And Phenomenology,''


\bibitem{Hartle}
J.~B.~Hartle,
``Space-time quantum mechanics and the quantum mechanics of space-time,''
arXiv:gr-qc/9304006.


\bibitem{Hawking}
S. Hawking, {\em  Particle Creation By Black Holes}
 Commun.Math.Phys. 43 (1975) 199

\bibitem{Hawking euc}
G.~W.~.~Gibbons and S.~W.~.~Hawking,
``Euclidean Quantum Gravity,''


\bibitem{hooft}
G.~'t Hooft and M.~J.~G.~Veltman,
``One Loop Divergencies In The Theory Of Gravitation,''
Annales Poincare Phys.\ Theor.\ A {\bf 20} (1974) 69.




\bibitem{Horowitz}
G.~T.~Horowitz,
``Quantum gravity at the turn of the millennium,''
arXiv:gr-qc/0011089.

 
\bibitem{Horowitz1}
G. Horowitz,''Exactly Soluble Diffeomorphism Invariant Theories'',
 Annals Phys. 205 (1991) 130


\bibitem{Hull}
C.~M.~Hull and P.~K.~Townsend,
``Unity of superstring dualities,''
Nucl.\ Phys.\ B { 438} (1995) 109
[arXiv:hep-th/9410167].



\bibitem{Immirzi}
G.~Immirzi,
``Quantum gravity and Regge calculus,''
Nucl.\ Phys.\ Proc.\ Suppl.\  { 57} (1997) 65
[arXiv:gr-qc/9701052].

\bibitem{Isham}
C.~J.~Isham,
``Structural issues in quantum gravity,''
arXiv:gr-qc/9510063.

\bibitem{Julve}
J.~Julve and M.~Tonin,
``Quantum Gravity With Higher Derivative Terms,''
Nuovo Cim.\ B {\bf 46} (1978) 137.




\bibitem{Kawai}
H.~Kawai, D.~C.~Lewellen and S.~H.~H.~Tye,
``A Relation Between Tree Amplitudes Of Closed And Open Strings,''
Nucl.\ Phys.\ B {\bf 269} (1986) 1.






\bibitem{Kodama}
H.~Kodama,
``Holomorphic Wave Function Of The Universe,''
Phys.\ Rev.\ D { 42} (1990) 2548.



\bibitem{Luscher}
M.~Luscher, R.~Narayanan, P.~Weisz and U.~Wolff,
``The Schrodinger functional: A Renormalizable probe for nonAbelian gauge theories,''
Nucl.\ Phys.\ B { 384} (1992) 168
[arXiv:hep-lat/9207009].



\bibitem{MM}
Y.~M.~Makeenko and A.~A.~Migdal,
``Exact Equation For The Loop Average In Multicolor QCD,''
Phys.\ Lett.\ B { 88} (1979) 135
[Erratum-ibid.\ B { 89} (1980) 437].





\bibitem{Maldacena}J. Maldacena, 
                   {\em The large N limit of superconformal field
                    theories and supergravity},
                   Adv.Theor.Math.Phys.2:231-252,1998, 
                    {\tt hep-th/9711200}. 

\bibitem{Montonen}
C.~Montonen and D.~I.~Olive,
``Magnetic Monopoles As Gauge Particles?,''
Phys.\ Lett.\ B {\bf 72} (1977) 117.
   

\bibitem{Myers}
R.~C.~Myers and M.~Pospelov,
``Experimental challenges for quantum gravity,''
arXiv:hep-ph/0301124.


\bibitem{Osborn}
H.~Osborn,
``Topological Charges For N=4 Supersymmetric Gauge Theories And Monopoles Of
Spin 1,''
Phys.\ Lett.\ B {\bf 83} (1979) 321.





\bibitem{penrose}
R. Penrose, {\em Angular momentum: an approach to combinatorial spacetime},in 
{\em Quantum theory and beyond}, T. Bastin ed.(Cambridge University Press, 1971)


\bibitem{Perelman}G. Perelman,
``The entropy formula for the Ricci flow and its geometric applications'' 


\bibitem{Pfeiffer}
H.~Pfeiffer,
arXiv:gr-qc/0404088.


\bibitem{Polchinski}
J.~Polchinski,
``Dirichlet-Branes and Ramond-Ramond Charges,''
Phys.\ Rev.\ Lett.\  { 75} (1995) 4724
[arXiv:hep-th/9510017].

\bibitem{Polchinskis}
J.~Polchinski,
``String Theory. Vol. 1: An Introduction To The Bosonic String,''
\\
``String Theory. Vol. 2: Superstring Theory And Beyond,''



\bibitem{Polyakov}
A.~M.~Polyakov,
``Quantum Geometry Of Bosonic Strings,''
Phys.\ Lett.\ B { 103} (1981) 207.




\bibitem{Rovellir}
C.~Rovelli,
``Loop quantum gravity,''
Living Rev.\ Rel.\  { 1} (1998) 1
[arXiv:gr-qc/9710008].



\bibitem{Rovelli}
C. Rovelli 
{\em  Loop Quantum Gravity} gr-qc/9606088 [=Phys.Lett. B380 (1996) 257]\\
``The Immirzi Parameter In Quantum General Relativity'',
 gr-qc/9505012 [=Phys.Lett. B360 (1995) 7]


\bibitem{RS}
C.~Rovelli and L.~Smolin,
``Spin networks and quantum gravity,''
Phys.\ Rev.\ D { 52} (1995) 5743
[arXiv:gr-qc/9505006].

\bibitem{RSS}
C.~Rovelli and L.~Smolin,
``Discreteness of area and volume in quantum gravity,''
Nucl.\ Phys.\ B { 442} (1995) 593
[Erratum-ibid.\ B { 456} (1995) 753]
[arXiv:gr-qc/9411005].


\bibitem{Scherk}
J.~Scherk and J.~H.~Schwarz,
``Dual Models For Nonhadrons,''
Nucl.\ Phys.\ B { 81} (1974) 118.









\bibitem{Smolin}
L.~Smolin,
``How far are we from the quantum theory of gravity?,''
arXiv:hep-th/0303185.

\bibitem{Smolinc}
L.~Smolin,
``Quantum gravity with a positive cosmological constant,''
arXiv:hep-th/0209079.



\bibitem{Stelle}
K.~S.~Stelle,
``Renormalization Of Higher Derivative Quantum Gravity,''
Phys.\ Rev.\ D {\bf 16} (1977) 953.





\bibitem{Strominger}
A.~Strominger and C.~Vafa,
``Microscopic Origin of the Bekenstein-Hawking Entropy,''
Phys.\ Lett.\ B { 379} (1996) 99
[arXiv:hep-th/9601029].


\bibitem{Susskind}
L.~Susskind,
``The World as a hologram,''
J.\ Math.\ Phys.\  { 36} (1995) 6377
[arXiv:hep-th/9409089].

\bibitem{'tHooft}
G.~'t Hooft,
``Dimensional Reduction In Quantum Gravity,''
arXiv:gr-qc/9310026.



\bibitem{thv}
G.~'t Hooft and M.~J.~Veltman,
``One Loop Divergencies In The Theory Of Gravitation,''
Annales Poincare Phys.\ Theor.\ A { 20} (1974) 69.

\bibitem{Thiemannr}
T.~Thiemann,
``Introduction to modern canonical quantum general relativity,''
arXiv:gr-qc/0110034.




\bibitem{Thiemann}
T.~Thiemann,
``Anomaly-free formulation of non-perturbative, four-dimensional  Lorentzian quantum gravity,''
Phys.\ Lett.\ B { 380} (1996) 257
[arXiv:gr-qc/9606088].\\
``Gauge field theory coherent states (GCS). I: General properties,''
Class.\ Quant.\ Grav.\  { 18} (2001) 2025
[arXiv:hep-th/0005233].




\bibitem{Turaev}
V.~G.~Turaev and O.~Y.~Viro,
``State Sum Invariants Of 3 Manifolds And Quantum 6j Symbols,''
Topology { 31} (1992) 865.















\bibitem{Witten} E.~Witten,
                 {\sl Anti de Sitter Space and Holography},
                 {\tt hep-th/9802150};\\
 {\sl Anti-de Sitter space,
                   thermal phase transition and confinement in gauge
                   theories}
                 {\tt hep-th/9803131}.
                
\bibitem{Witten3}
E.~Witten,
``(2+1)-Dimensional Gravity As An Exactly Soluble System,''
Nucl.\ Phys.\ B { 311} (1988) 46.

\bibitem{Wittenk}
E.~Witten,
``A note on the Chern-Simons and Kodama wavefunctions,''
arXiv:gr-qc/0306083.

                 
\end{thebibliography}
\end{document}